\documentclass[12pt,preprint]{aastex}
\clearpage

\shorttitle{Mass Transfer in RY Per}
\shortauthors{Barai et al.}

\begin{document}

\received{2003 September 26}
\accepted{}

\title{Mass and Angular Momentum Transfer \\ in the Massive Algol Binary RY Persei}

\author{P. Barai, 
        D. R. Gies\altaffilmark{1}, 
        E. Choi,
        V. Das,
        R. Deo,
        W. Huang\altaffilmark{1},
        K. Marshall, 
        M. V. McSwain\altaffilmark{1},
        C. Ogden, 
        M. A. Osterman, 
        R. L. Riddle\altaffilmark{1,2}, 
        J. L. Seymour, Jr., 
        D. W. Wingert\altaffilmark{1}}

\affil{Center for High Angular Resolution Astronomy\\
Department of Physics and Astronomy \\
Georgia State University, University Plaza, Atlanta, GA  30303-3083\\
Electronic mail: barai@chara.gsu.edu, gies@chara.gsu.edu, echoi@chara.gsu.edu, 
das@chara.gsu.edu, deo@chara.gsu.edu, huang@chara.gsu.edu, 
marshall@chara.gsu.edu, mcswain@chara.gsu.edu, ogden@chara-array.org,
osterman@chara.gsu.edu, riddle@iastate.edu, seymour@chara.gsu.edu, wingert@chara.gsu.edu}

\altaffiltext{1}{Visiting Astronomer, Kitt Peak National Observatory,
National Optical Astronomy Observatories, operated by the Association
of Universities for Research in Astronomy, Inc., under contract with
the National Science Foundation.}

\altaffiltext{2}{Current address: Department of Physics and Astronomy,
Iowa State University, Ames, IA 50011}

\author{A. B. Kaye\altaffilmark{1,3}}
\affil{
ITT Industries, Advanced Engineering and Sciences Division,
1550 Wyoming Blvd. SE, 
Kirtland Air Force Base, New Mexico 87117 \\
Electronic mail: anthony.kaye@kirtland.af.mil}

\altaffiltext{3}{Dr.\ Kaye is on contract to the USAF/XON Nuclear Weapons 
and Counterproliferation Agency.}

\author{G. J. Peters\altaffilmark{4}}
\affil{Space Sciences Center, University of Southern California,\\
University Park, Los Angeles, CA 90089 \\
Electronic mail: gjpeters@mucen.usc.edu}

\altaffiltext{4}{Visiting Astronomer, International 
Ultraviolet Explorer Satellite.}

\slugcomment{Submitted to ApJ}
\paperid{59050}


\begin{abstract}
We present an investigation of H$\alpha$ emission line variations
observed in the massive Algol binary, RY~Per. 
We give new radial velocity data for the secondary based upon our
optical spectra and for the primary based upon high dispersion UV spectra. 
We present revised orbital elements and an estimate of the
primary's projected rotational velocity (which indicates that 
the primary is rotating $7\times$ faster than synchronous). 
We use a Doppler tomography algorithm to reconstruct the 
individual primary and secondary spectra in the region of H$\alpha$, 
and we subtract the latter from each of our observations to 
obtain profiles of the primary and its disk alone. 
Our H$\alpha$ observations of RY~Per show that the mass gaining 
primary is surrounded by a persistent but time variable accretion disk. 
The profile that is observed outside-of-eclipse has weak, double-peaked 
emission flanking a deep central absorption, and we find that these 
properties can be reproduced by a disk model that includes 
the absorption of photospheric light by the band of the disk 
seen in projection against the face of the star.  
We developed a new method to reconstruct the disk surface 
density distribution from the ensemble of H$\alpha$ profiles 
observed around the orbit, and this method accounts for the 
effects of disk occultation by the stellar components, 
the obscuration of the primary by the disk, and flux 
contributions from optically thick disk elements.  
The resulting surface density distribution is elongated 
along the axis joining the stars, in the same way as seen 
in hydrodynamical simulations of gas flows that strike the 
mass gainer near trailing edge of the star.  This type 
of gas stream configuration is optimal for the transfer of 
angular momentum, and we show that rapid rotation is found 
in other Algols that have passed through a similar stage.   
\end{abstract}
 


\section{Introduction}                              

Algol-type binary systems consist of a hot, B-A main sequence primary star 
and a cool, less massive F-K giant or subgiant secondary companion 
in a close orbit \citep{pet89,pet01}. 
They appear to have formed from previously detached binaries in which the 
originally more massive star (now the secondary) evolves off the main sequence, 
expands to its Roche lobe limit, and begins to transfer mass and 
angular momentum to its companion (the primary).  
The gas stream from the donor to the gainer follows a path that 
directly strikes the star in short period systems while in longer period systems 
($P>6$~d) the stream misses the star and forms an accretion disk. 
The circumstellar gas is observed in UV and optical emission lines, 
and the H$\alpha$ line in particular has served as an important 
diagnostic of the outflow \citep{ric99,ric01}.  
H$\alpha$ profiles corrected for photospheric absorption show single and/or double-peaked 
emission features at most orbital phases.  \citet{ric99} have made Doppler 
tomographic reconstructions of the H$\alpha$ emission in many systems, and 
these tomograms are important diagnostics in searching for evidence of the
gas streams and disks. 

RY~Persei (HD~17034 = HIP~12891 = BD$+47^\circ 692$) is a remarkable 
example of one of the most massive Algol binaries.   
In a seminal paper \citet{ols97} determined accurate values of the 
masses and radii of this totally eclipsing system ($P=6.86$~d): 
$M_P/M_\odot = 6.25\pm 0.16$ and $R_P/R_\odot = 4.06\pm 0.14$ for the B4:~V gainer and 
$M_S/M_\odot = 1.60\pm 0.10$ and $R_S/R_\odot = 8.10\pm 0.17$ for the F7:~II-III donor star. 
\citet{ols97} discuss the ample evidence for ongoing mass transfer and 
circumstellar gas, and they point out that the gainer is rotating approximately 
$10\times$ faster than the orbital synchronous rate.  
Thus, RY~Per offers us a key opportunity to investigate the process of mass transfer
and the resulting spin up of the gainer in a system with well known properties. 
We are particularly interested in learning whether or not this process 
of spin up is a possible explanation for the origin of the Be and other rapidly rotating 
massive stars \citep{gie98,gie01}.   

Here we present the results of an investigation of new  
high quality H$\alpha$ spectroscopy of RY~Per obtained with the 
Kitt Peak National Observatory (KPNO) Coude Feed Telescope. 
In \S2 we describe the observations (which include two intensive series of 
spectra obtained in 1999 October and 2000 October during total eclipses of the primary)
and new radial velocity data for the secondary.  We also present radial velocities
for the primary derived from high dispersion UV spectroscopy made with 
the {\it International Ultraviolet Explorer Satellite (IUE)}. 
We present revised orbital elements in \S3 and discuss the 
projected rotational velocity of the primary in \S4. 
We show the results of a Doppler tomographic reconstruction of the 
stellar component spectra in \S5, where we also discuss the removal 
of the secondary's spectrum from the composite spectra. 
We present in \S6 a disk model of the time- and orbital phase-averaged H$\alpha$ 
profile using the method of \citet{hum00} that is applicable to Be ``shell'' stars. 
Our study indicates, however, that there is evidence of non-axisymmetric 
structure in the disk of RY~Per, and in \S7 we present a new method to 
reconstruct the disk surface density distribution from the H$\alpha$ observations. 
We summarize our results in \S8 and discuss how RY~Per is representative 
of Algol systems at a stage where the angular momentum transfer efficiency peaks. 


\section{Observations and Radial Velocity Measurements} 

The optical spectra were obtained with the KPNO
0.9-m Coude Feed Telescope during four runs between 1999 October and
2000 October.   A summary of the different observing runs is
given Table~1 which lists the beginning and ending UT
dates of observation, grating, filter, 
spectral resolving power ($\lambda/\delta\lambda$), and 
wavelength range recorded.  
In the first two runs we used
the short collimator, grating BL181, and camera 5 with a Ford $3072
\times 1024$ CCD (F3KB) as the detector.  
During the third and fourth runs we used the same camera and detector 
but changed to the long collimator with higher dispersion gratings 
(gratings A and B in the third and fourth runs, respectively). 
Exposure times were generally 30 minutes, which produced a 
signal-to-noise ratio of $S/N\approx 300$ in the lower dispersion,
out-of-eclipse spectra.  
We also observed with each configuration the rapidly rotating A-type star,
$\zeta$~Aql, which we used for removal of atmospheric water vapor lines.
Each set of observations was accompanied by numerous bias and flat field
frames, and wavelength calibration frames of a Th~Ar comparison lamp 
were obtained at approximately 90 minute intervals each night. 

\placetable{tab1}      

The spectra were extracted and calibrated
using standard routines in IRAF\footnote{IRAF is distributed by the
National Optical Astronomy Observatories, which is operated by
the Association of Universities for Research in Astronomy, Inc.,
under cooperative agreement with the National Science Foundation.}.
All the spectra were rectified to a unit continuum by fitting
line-free regions.  The removal of atmospheric lines was done by
creating a library of $\zeta$~Aql spectra from each run, removing
the broad stellar features from these, and then dividing each target
spectrum by the modified atmospheric spectrum that most closely
matched the target spectrum in a selected region dominated by
atmospheric absorptions.  The spectra from each run were then
transformed to a common heliocentric wavelength grid.

We measured radial velocities for the secondary by cross correlating 
each spectrum with a mid-eclipse spectrum (from HJD~2,451,463.804) 
that is dominated by the secondary's flux.   We omitted from the 
calculation those spectral regions containing strong interstellar and 
primary star lines.   Radial velocities were obtained by fitting a 
parabola to the cross correlation function extremum.  
Note that there is no problem in using a lower resolution spectral 
template (from the first run) in forming cross correlation functions
with higher resolution spectra (from the third and fourth runs), since 
the degraded resolution simply broadens the cross correlation function
but does not affect its central position. 
We transformed these relative velocities to an absolute scale by measuring 
the cross correlation shift between the mid-eclipse reference spectrum 
and a spectrum we made of HD~216228 (K0~III) which 
has a radial velocity of $-14.2\pm 0.7$ km~s$^{-1}$ \citep{bar00}.
We observed this radial velocity standard star in all but the 
third run, and cross correlation measurements indicate that any 
systematic errors between the runs are negligible ($<1.4$ km~s$^{-1}$).  
Our final results are presented in Table 2, which lists
the heliocentric Julian date of mid-exposure, the spectroscopic 
orbital phase (\S3), the radial velocity, and the observed minus
calculated radial velocity residual (\S3).

\placetable{tab2}                

The \ion{He}{1} $\lambda\lambda 5876,6678$ lines of the primary appear
to be partially blended with lines from the secondary and circumstellar components,
so they are not suitable for measuring the orbital motion of the primary. 
Instead we measured radial velocities of the primary in the spectral range 
1240 -- 1850 \AA ~recorded in {\it International Ultraviolet Explorer Satellite} 
high dispersion, short wavelength prime camera spectra.   
The secondary has negligible flux in this spectral 
region.   We obtained radial velocities by cross correlating each spectrum 
with the narrow-lined spectrum of $\iota$~Her (B3~IV), which has a radial 
velocity of $-20.0$ km~s$^{-1}$ \citep{abt78}.  Here we omitted rom the 
calculation any spectral regions containing strong interstellar and circumstellar 
lines.   Our measured absolute radial velocities for the primary are 
collected in Table~3 (in the same format as Table 2).

\placetable{tab3}                
 

\section{Revised Orbital Elements}                  

The last spectroscopic orbital solution was made by \citet{pop89}, 
and we decided it was worthwhile to check the period and orbital 
elements with the new data.  We first determined times of primary 
mid-eclipse from our intensive spectroscopic coverage of two eclipses 
by finding the times of maximum secondary line depth
(HJD $2451463.817 \pm 0.020$ and $2451820.731 \pm 0.020$). 
Both of these times are consistent with the primary eclipse
ephemeris of \citet{ols97} and of \citet{kre01}, so we adopted
in our radial velocity analysis the orbital period assumed by 
\citet{ols97}, $P = 6.863569$~d, which is based on the analysis  
by \citet{van86} of photometry by \citet{pop77}.  

The radial velocity curve of the secondary can be determined much 
more accurately than that of the primary because the measurement 
errors are smaller (more and narrower lines) and because we have more measurements. 
We determined orbital elements for both stars using the non-linear 
least-squares fitting code of \citet{mor74}. 
We first solved for the five orbital elements for the secondary: eccentricity ($e$), 
longitude of periastron ($\omega_S$), semiamplitude ($K_S$), systemic velocity ($V_S$),
and time of periastron ($T_S$).   
The fit of the primary radial velocities was made by fixing $e$, 
$\omega_P = \omega_S + 180^\circ$, and $P$ using the secondary's solution. 
The resulting orbital elements are 
listed in Table~4 (together with those from \citet{pop89}) and 
the radial velocity curve and observations are shown in Figure~1.
The difference in the systemic velocity between the solutions for 
the primary and secondary is probably not significant given the very different
ways in which the radial velocities were measured in the UV and optical spectra. 
The masses and semimajor axis were calculated using $i=83\fdg0$ \citep{ols97}.

Our results are in good agreement with those of \citet{pop89}. 
The main difference is our finding that the eccentricity is 
significantly different from zero, and this is somewhat surprising 
given that most Roche lobe overflow systems have circular orbits.  
However, Algol itself has a small but non-zero eccentricity 
\citep{hil71}, which may be related to the presence of 
a distant third star \citep{don95}. 
A statistical test of the significance of a non-zero 
eccentricity in RY~Per was made following \citet{luc71} by evaluating the 
probability that the reduction in the error of the fit through 
the addition of two parameters in an elliptical solution 
($e, \omega$) could have exceeded that obtained, i.e., 
\begin{displaymath}
p = (R_e / R_c)^\beta
\end{displaymath}
where $R_e$ is the sum of the squares of the residuals for 
an elliptical fit, $R_c$ is the same for a circular fit,
$\beta = (N-M)/2$, $N$ is the number of observations, and 
$M$ is the number of fitting parameters ($M=5$ since 
the period $P$ was set independently). 
We found that the r.m.s.\ of the fit dropped from 
5.1 km~s$^{-1}$ for a circular fit to 
3.4 km~s$^{-1}$ for an elliptical fit, and 
the probability that such an improvement would occur by 
random errors is $p=10^{-8}$.   
For the remainder of this paper we will refer 
to the photometric convention for orbital phase (0 at primary 
eclipse), $\phi_{\rm phot} = \phi_{\rm spec}-0.541$. 

\placetable{tab5}      

\begin{figure}
\rotatebox{90}{
\epsscale{0.7}
\plotone{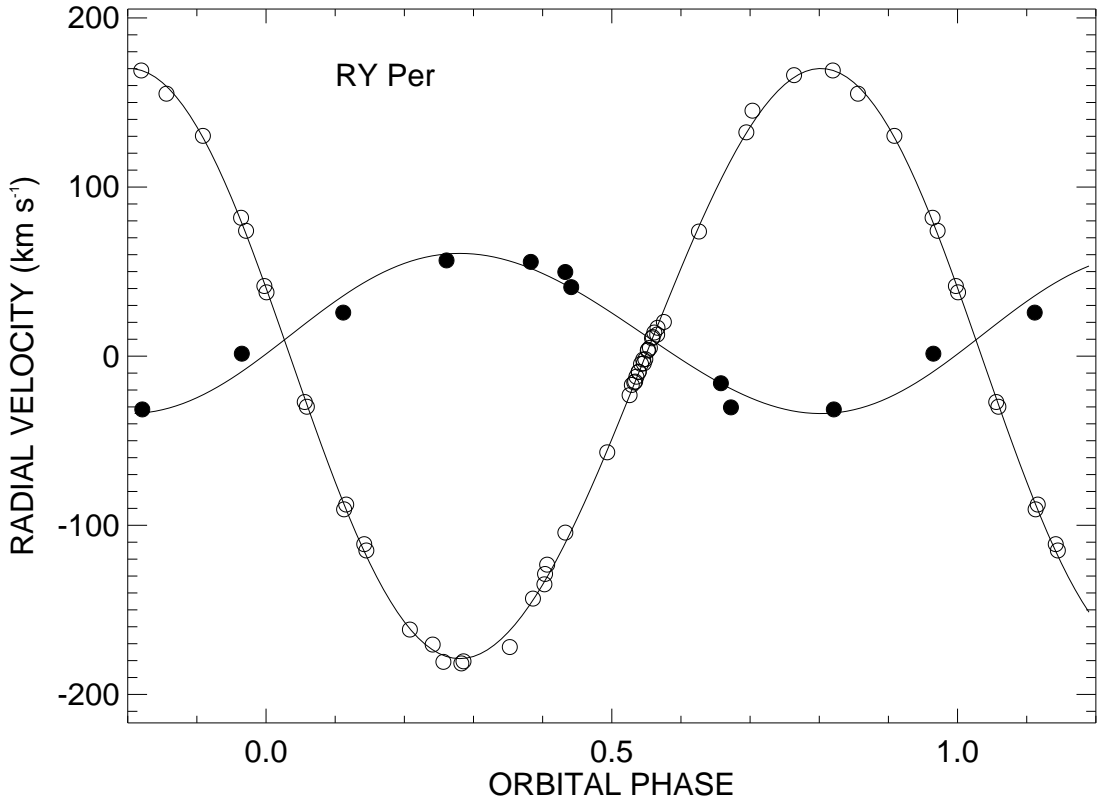}}
\caption{The fitted radial velocity curves ({\it solid lines}) for RY~Per 
together with the observed data for the secondary ({\it open circles}) 
and the primary star ({\it filled circles}).  Phase 0.0 corresponds to 
periastron.}
\label{fig1}
\end{figure}

\placefigure{fig1}     


\section{Projected Rotational Velocity of the Primary} 

We measured the projected rotational broadening of the primary's
UV lines following the method of \citet{how97}.   This method uses 
the width of the cross correlation function (ccf) discussed in 
\S2 as a measure of the photospheric line broadening 
(which is dominated by rotational broadening for the UV lines 
included in the construction of the ccf).   
First the UV reference spectrum of $\iota$~Her was cross correlated with itself 
to produce a sharp autocorrelation whose width measures the thermal 
broadening of the lines and any systematic effects due to line blending. 
We then assumed that the observed ccf was a convolution of the 
$\iota$~Her autocorrelation function (for $V\sin i =0$; \citet{abt02}) with a simple rotational 
broadening function based upon a linear limb darkening coefficient \citep{gra92}.
The limb darkening coefficient was taken from the tables of \citet{wad85} 
using the stellar effective temperature and gravity given by \citet{ols97}. 
The best fit of the broadened autocorrelation with the observed ccf occurred for
$V\sin i =  213  \pm 10 $ km~s$^{-1}$, which is in good agreement with the value
of $V\sin i =  212  \pm 7 $ km~s$^{-1}$ found by \citet{etz93} from optical lines. 
This implies that the primary is spinning $(7.2\pm 0.3)\times$ faster 
than the synchronous rate (given the stellar radius and system 
inclination from \citet{ols97}). 


\section{Spectral Components and H$\alpha$ Emission} 

Our primary focus in this paper is the H$\alpha$ emission flux 
that is formed in the vicinity of the hot primary star.   However, the
cool secondary star's spectrum also displays many line features in 
this same spectral region (including H$\alpha$ in absorption).  
Thus, we need to remove the secondary star's spectral contribution 
from the observed spectra in order to study the H$\alpha$ contributions 
from the primary star and its circumstellar gas.   

There are many approaches to finding a suitable representation 
of the secondary's spectrum for this purpose (i.e., using a spectrum 
of a comparable field F-type giant star or using a spectrum of RY~Per 
obtained near mid-eclipse), but we decided that best method was to
isolate the secondary's spectrum using our composite spectra for 
orbital phases outside of the eclipses.  We made a reconstruction of 
the individual primary and secondary spectra from our composite 
spectra using the Doppler tomography algorithm discussed by \citet{bag94}, 
which we have used to great effect in studies of other massive binaries
\citep{har02,gie02,pen02}.   The algorithm assumes that each observed 
composite spectrum is a linear combination of spectral components with 
known radial velocity curves (\S3) and flux ratios.  In the case of RY~Per, 
the flux ratio varies significantly across the wavelength range 
recorded in our spectra because of the large differences in the 
temperatures and spectral flux distributions of the stellar components. 
We modified the tomography code to use a flux ratio dependent on wavelength, 
and we calculated the appropriate flux ratio based upon the $V$ band 
magnitude difference derived by \citet{ols97} and model flux 
distributions from \citet{kur94} (for assumed effective temperatures 
of $T_{\rm eff} = 18000$ and 6250~K and gravities of $\log g = 4.02$ 
and 2.83 for the primary and secondary, respectively; \citet{ols97}). 
We also introduced a fourth-order polynomial fit of line-free regions
during each iteration of the tomography algorithm in order to avoid 
unrealistic variations in continuum placement.   The resulting 
tomographic reconstructions of the primary and secondary spectra 
are illustrated in Figure~2.  

\begin{figure}
\rotatebox{90}{
\epsscale{0.7}
\plotone{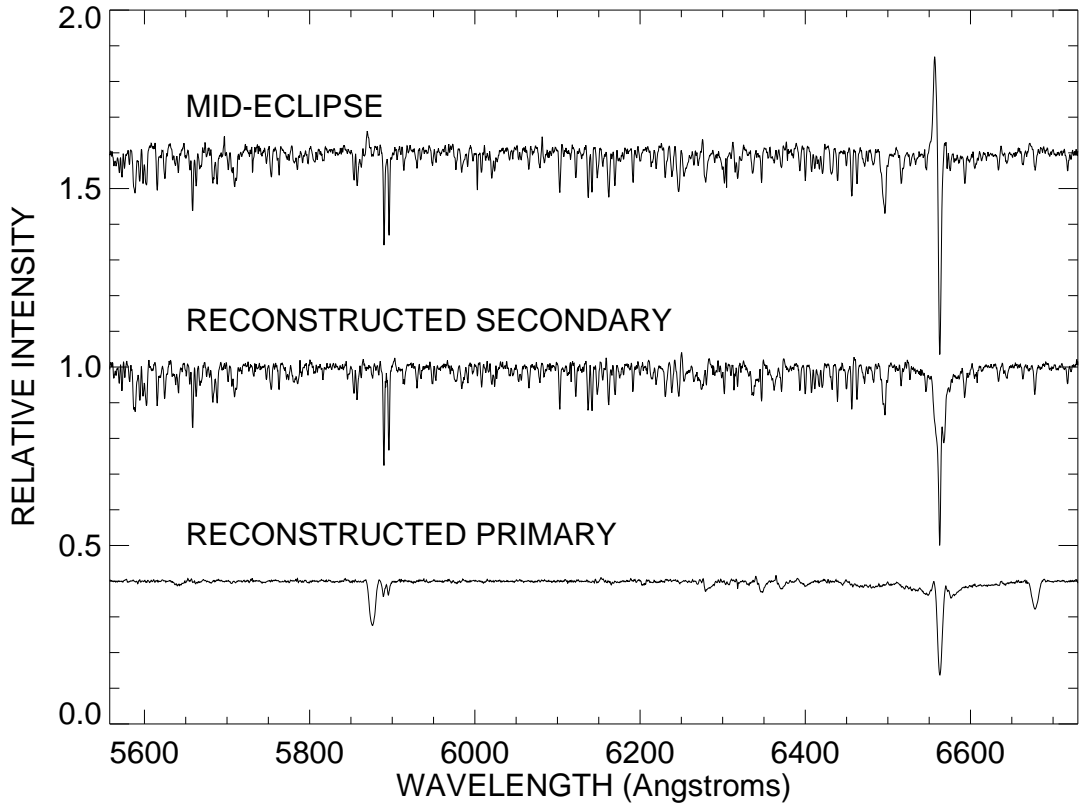}}
\caption{Tomographically reconstructed spectra of the primary ({\it bottom})
and secondary ({\it middle}) based upon outside-of-eclipse spectra
and rectified to a unit continuum. 
The upper plot shows a single spectrum obtained at mid-eclipse when the flux 
is entirely due to the secondary (except for disk emission at H$\alpha$).}
\label{fig2}
\end{figure}

\placefigure{fig2}     

The primary star's spectrum shows the important H$\alpha$ and 
\ion{He}{1} $\lambda\lambda 5876, 6678$ lines found in B-type spectra. 
The H$\alpha$ profile appears to be partially filled-in with 
weak, double-peaked emission that is normally associated with a
disk \citep{ric99,ric01}.  The primary's spectrum also shows weak 
and narrow \ion{Na}{1} $\lambda\lambda 5890, 5896$ lines, which we 
suspect result from incomplete removal of the strong interstellar 
components of Na~D.   The secondary spectrum displays a rich 
collection of metallic lines and strong Na~D components.  
It also has a deep and narrow H$\alpha$ absorption that agrees in 
shape with the predicted profile from the models of \citet{kur94} 
for the secondary's effective temperature and gravity.   We also 
show in Figure~2 a single spectrum obtained near mid-eclipse  
(made at HJD 2,451,463.850) when the flux should be totally 
due to the secondary (except for some disk emission at H$\alpha$). 
The excellent agreement between the mid-eclipse spectrum 
and the reconstructed secondary spectrum is an independent 
verification of the reliability of the tomography algorithm. 

We removed the secondary's spectrum from the observed 
composite spectrum and renormalized the difference relative 
to the primary's continuum flux outside-of-eclipse.  
Let $P$, $S$, and $C$ represent the continuum rectified versions of
the spectra of the primary, secondary, and their combination, respectively.
We express the phase variable continuum fluxes as $F_P$ and $F_S$ for the 
primary and secondary, which are given in units of the primary continuum
flux outside-of-eclipse, $F_P^o$.  Then the isolated primary 
spectrum in units of its flux outside-of-eclipse is given by 
\begin{equation}
{F_P \over F_P^o} P = {{F_P+F_S} \over F_P^o}~C - {F_S \over F_P^o}~S.
\end{equation}
We estimated the flux ratio $F_P / F_S$ in individual spectra 
by measuring the dilution of the secondary's line depths due to the 
additional continuum flux of the primary, 
\begin{equation}
d_{\rm observed} = d_{\rm actual} (1 + {F_P \over F_S})^{-1}, 
\end{equation}
where $d_{\rm actual}$ refers to line depths in the reconstructed 
secondary spectrum.   For spectra obtained outside-of-eclipse, 
we subtracted out the reconstructed secondary spectrum $S$ by adopting
$F_P / F_P^o = 1$ and using the observed secondary 
line depths to estimate $F_S / F_P^o$ (eq.~2).  On the other hand, 
for spectra obtained during eclipses, we used the secondary line depths
to establish the flux scale (since the secondary flux is approximately 
constant during the eclipse of the primary) so that the primary's flux is
given by 
\begin{equation}
{F_P \over F_P^o} = {F_P \over F_S} ~ {F_S \over F_P^o}.
\end{equation}
We again found $F_P / F_S$ from the observed secondary line depths
and we used the default (outside-of-eclipse) flux ratio adopted in the 
tomographic reconstruction to set $F_S / F_P^o$.  The difference 
spectra for the primary eclipse phases are then referenced to a
constant flux scale to facilitate inspection of the disk emission component. 

The secondary-subtracted spectra are illustrated as a function of photometric
orbital phase (primary eclipse at $\phi = 0.0$) in Figures 3, 4, and 5.   
Figure~3  shows the out-of-eclipse 
sample in the radial velocity frame of the primary star.  Each spectrum 
is positioned along the $y$-axis so that the primary's rectified continuum 
flux is set equal to the orbital phase of observation.   Most of these
spectra are characterized by double-peaked emission with a strong central 
absorption, but there are significant variations that are both related and 
unrelated to orbital phase.  Most of the spectra obtained in the phase range
0.8 -- 1.0 show evidence of strong, red-shifted absorption (also seen in the
\ion{He}{1} lines).  This phase, just prior to primary eclipse, is when the 
terminal point of gas stream from the secondary to the primary is seen projected 
against the photosphere of the primary star (discussed below in \S7), and 
the extra absorption may be due to atmospheric extension at the stream -- 
star impact site. 

\begin{figure}
\plotone{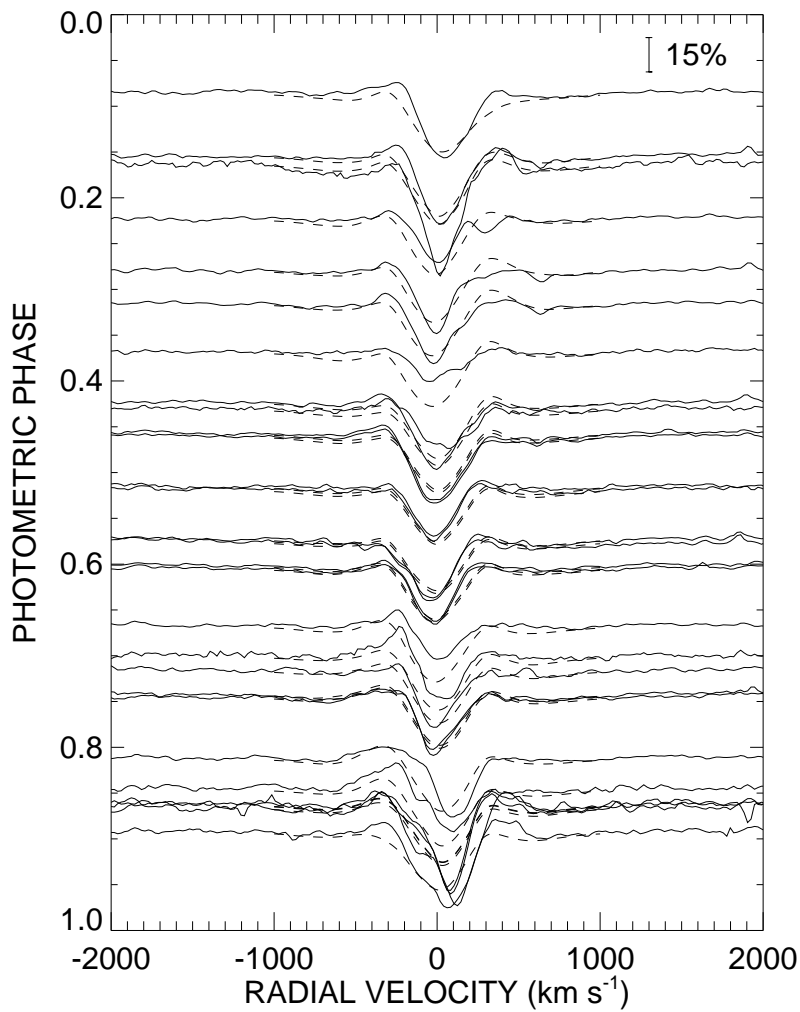}
\caption{Secondary subtracted H$\alpha$ profiles for the out-of-eclipse sample.
The spectra are plotted in the velocity frame of the primary star.
Each spectrum is aligned so that the continuum level is set at the orbital 
phase of observation (phase 0.0 corresponds to primary eclipse).  
The bar at the upper right gives the spectral intensity
scale relative to a primary continuum flux of unity.  The dashed lines represent
model spectra derived from the reconstruction of the disk surface density (\S7).}
\label{fig3}
\end{figure}

\placefigure{fig3}     

\begin{figure}
\plotone{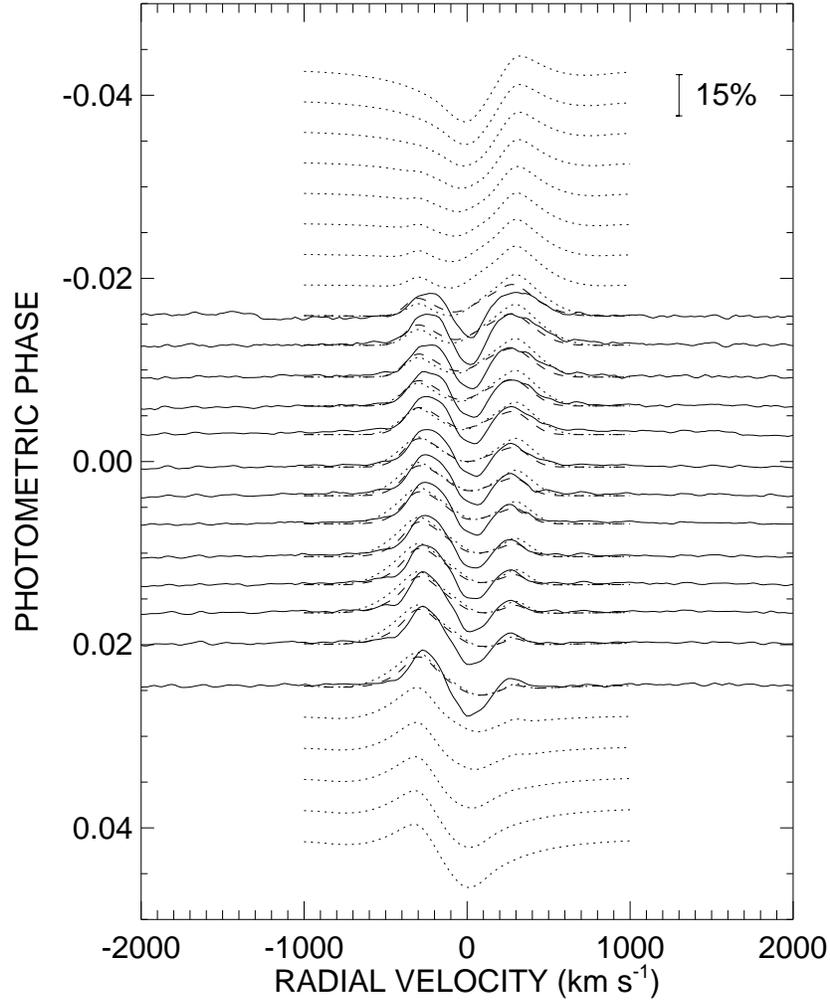}
\caption{Secondary subtracted H$\alpha$ profiles for the eclipse observed 
on 1999 October 12 (in a format similar to  Fig.~3).  The dotted lines show model 
predictions for eclipses of an axisymmetric disk (\S6), and these are extended 
before and after the observed sequence to show the predicted behavior beyond the
times of mid-eclipse. The dashed lines represent
model spectra derived from the reconstruction of the disk surface density (\S7).}
\label{fig4}
\end{figure}

\placefigure{fig4}     

\begin{figure}
\plotone{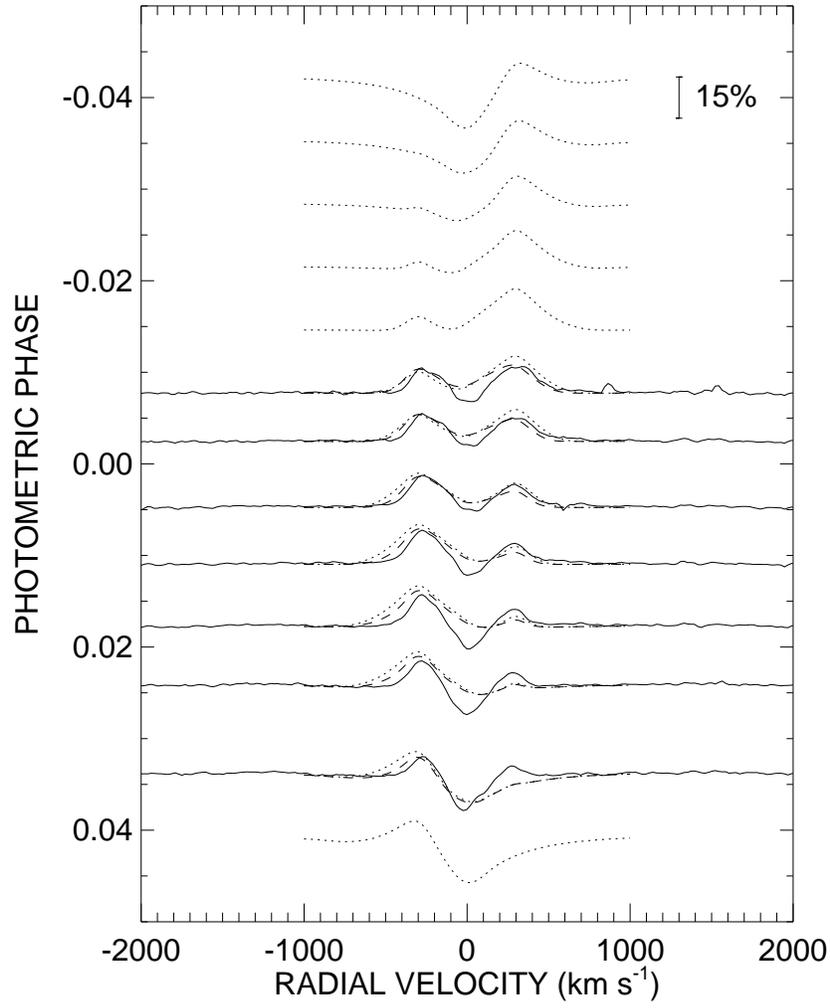}
\caption{Secondary subtracted H$\alpha$ profiles for the eclipse observed 
on 2000 October 03 (in the same format as Fig.~4). }
\label{fig5}
\end{figure}

\placefigure{fig5}     

Figures 4 and 5 show our intensive coverage of the H$\alpha$ profiles 
during the two eclipses of 1999 October 12 and 2000 October 03, respectively.
Once again, the continuum level is aligned with the phase of observation, 
but for these eclipse phases the continuum flux is less than unity (and should be 
zero between phases $-0.007$ and $+0.007$ according to the light curve of
\citet{ols97}).  In both sets of eclipse spectra, we see that the disk 
emission is never fully eclipsed.  The relative strengths of the blue and 
red peaks reverse as the secondary first occults the approaching and then 
the receding portions of the disk.  The central absorption remains strong 
throughout these sequences and in fact dips below zero intensity relative 
to the far wings in a number of the spectra obtained around mid-eclipse in 
the 1999 October 12 observations.  The central absorption outside-of-eclipse
is probably due in large measure to the obscuration of the primary's photosphere 
by its circumstellar disk (\S6), but this cannot be the case around mid-eclipse 
since the primary is totally occulted by the secondary.  The simplest explanation 
is that there is additional, foreground circumstellar gas projected against the 
photosphere of the secondary around mid-eclipse that makes the secondary's 
H$\alpha$ absorption stronger than we assumed in subtracting the tomographically 
reconstructed secondary spectrum.   It is possible, for example, that the 
secondary occasionally loses mass from the L2 location into a circumbinary disk 
and that this counterstream causes the additional absorption seen around mid-eclipse. 

Both the eclipse and out-of-eclipse phase observations offer evidence 
of a persistent but temporally variable accretion disk around the primary star. 
In the following sections we will explore physical disk models that can 
explain the observed H$\alpha$ properties.  We first consider axisymmetric 
disk models (\S6) that can reproduce the orbital-average H$\alpha$ profile in 
the tomographic reconstruction of the primary, and then we explore non-axisymmetric
models (\S7) that can help explain the orbital-phase related variations in H$\alpha$. 


\section{Axisymmetric Disk Models}                  

The tomographically reconstructed H$\alpha$ profile of the primary 
is illustrated in Figure~6.  Similar double-peaked emission is found
in other Algol-type binaries with comparable periods and is generally 
interpreted as originating in an accretion disk surrounding the primary
\citep{ric99,ric01}.  The portions of the disk viewed against the sky 
produce the blue- and red-shifted emission peaks while the portion of 
the disk projected against the photosphere of the primary creates the 
central absorption feature.   A similar morphology is observed in 
Be ``shell'' stars, i.e., rapidly rotating, hot stars with a disk 
viewed nearly edge-on.  \citet{hum00} have developed a semi-analytical 
method to model the H$\alpha$ emission profiles of shell stars based 
upon the theory of shear broadening in the accretion disks of cataclysmic 
variables \citep{hor86}.   Their approach is valid for relatively 
low density disks that are mainly confined to the orbital plane and 
are viewed at inclinations somewhat less than $i=90^\circ$.   These 
conditions are met in the case of RY~Per, so we adopted their 
methods to produce disk profiles for comparison with our H$\alpha$ observations. 

\begin{figure}
\rotatebox{90}{
\epsscale{0.7}
\plotone{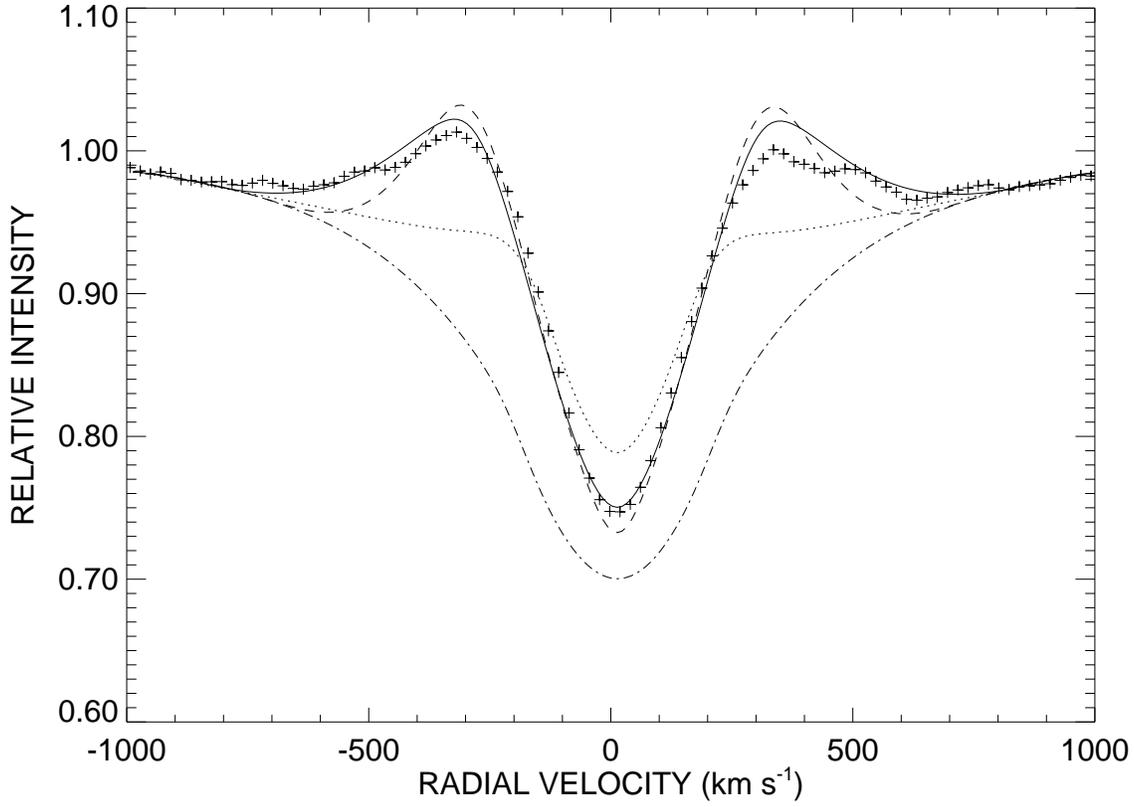}}
\caption{The tomographically reconstructed H$\alpha$ profile of the 
primary star ({\it plus signs}) together with several axisymmetric 
disk model profiles: 
$j=0.5$, $m=5.5$ ({\it solid line}); 
$j=0.5$, $m=2$ ({\it dashed line}); 
$j=1$, $m=6$ ({\it dotted line}). 
Also shown is the model rotationally broadened absorption profile 
for the photosphere of the primary ({\it dot-dashed line}).}
\label{fig6}
\end{figure}

\placefigure{fig6}     

The orbital dimensions and masses were set by our revised orbital elements (\S3)
and the orbital inclination found by \citet{ols97}, $i=83\fdg0 \pm 0\fdg3$.
The radius of the primary, $R_P = 4.06 R_\odot$, was taken from the analysis 
of \citet{ols97}.   In our numerical model the disk is assumed to be relatively thin 
and to extend from the photosphere to the Roche radius of the primary, 
$R_{\rm Roche} = 15.0 R_\odot$.   The actual disk area was subdivided into 131 radial ($R$) 
and 360 azimuthal ($\phi$) elements.   We artificially extended the grid inwards to $R=0$ 
and outwards to $R = R_P / \cos i = 33.3 R_\odot$, so that there are pseudo-disk 
elements that project onto the entire stellar photosphere, simplifying  
the numerical integration of flux from the photosphere.   The flux increment
from a given area element is given by (see eq.~14 in \citet{hum00}) 
\begin{equation}
F_\lambda = A~ [S^L (1 - e^{-\tau_\lambda}) + P_\lambda e^{-\tau_\lambda}]. 
\end{equation}
Here $A$ is the projected area of the element, 
$S^L$ is ratio of the H$\alpha$ line source function to the continuum 
flux of the primary, $\tau_\lambda$ is the optical depth within the element, 
and $P_\lambda$ is the stellar photospheric component at the projected 
position of the element.  \citet{hum00} advocate a disk temperature 
$T_d = {2\over 3} T_{\rm eff}$, which is the common assumption for Be 
star disks, so we adopted $T_d = 12000$~K for the disk in RY~Per. 
Then the line source function ratio is given by the ratio of the Planck functions, 
$S^L = B_\lambda(T_d)/ B_\lambda(T_{\rm eff}) = 0.456$. 
Following \citet{hum00} (their eq.~12), we check to 
see if an area element is projected against the sky or the star:
if projected against the sky, then we set $P_\lambda = 0$, but otherwise
$P_\lambda$ is set to be the limb darkened, intensity profile defined by $\mu$, 
the cosine of the angle between the photospheric normal and line of sight, 
and by the projected rotational velocity of photospheric element. These H$\alpha$ intensity 
profiles were calculated using the {\it Synspec} radiative transfer code and simple LTE model 
atmospheres from the {\it Tlusty} code\footnote{http://tlusty.gsfc.nasa.gov}
\citep{lan03}.  Note that we ignore rotational distortions and any associated 
gravity darkening in the primary star.   Elements in the pseudo-disk 
($R<R_P$ and $R>R_{\rm Roche}$) are assigned optical depth $\tau_\lambda = 0$, 
so that only photospheric contributions are included there.  The rotationally broadened 
photospheric flux profile we derive is shown in Figure~6. 

The optical depth of a disk element depends on the physical parameterization 
of the disk (see \citet{hum00} for details).  The disk azimuthal velocity is given by 
\begin{equation}
V_{\rm rot} = V(R_P) ~R^{-j}
\end{equation}
where the radial distance $R$ is given in units of stellar radius $R_P$.
The disk density is set by a radial power law ($n\propto R^{-m}$) and 
a Gaussian distribution normal to the disk characterized by the disk 
scale height $H(R)$ (see eq.~2--4 in \citet{hum00}).   These parameters 
are combined in the disk surface density, $\Sigma(R)$, i.e., the column density
integrated normal to the disk at a given position.  Then the optical depth is
given by (see eq.~7--9 in \citet{hum00})
\begin{equation}
\tau_\lambda = {{\pi e^2}\over{mc}} f \Sigma(R) {1\over{\cos i}} \lambda_0 
  {1\over{\sqrt{2\pi} \triangle V}} \exp [-{1\over 2}(\triangle V_r/\triangle V)^2]
\end{equation} 
where $\triangle V_r = [c (\lambda - \lambda_0) / \lambda_0] - V_r$ and
$V_r$ is the local radial velocity of the element. 
The width of this Gaussian distribution is defined by 
\begin{equation}
\triangle V = \sqrt{V_{\rm th}^2+V_{\rm sh}^2} 
\end{equation} 
where the thermal velocity broadening is $V_{\rm th} = 14.1$ km~s$^{-1}$ 
(for $T_d=12000$~K) and the shear broadening 
due to the Doppler velocity gradient along the line of sight is \citep{hor95}
\begin{equation}
V_{\rm sh} = -(j+1) {H\over R} V_{\rm rot} \sin\phi \cos\phi \sin i \tan i.
\end{equation} 
For Keplerian motion in the disk ($j=0.5$), $V_{\rm sh}$ ranges from 0 to 
100 km~s$^{-1}$ in our model (depending on the position of the element). 

Our sample H$\alpha$ model fits (smoothed to the instrumental 
broadening of the spectra) are plotted in 
Figure~6.  We explored two possible velocity laws (Keplerian, $j=0.5$, and
angular momentum conserving, $j=1$) and a range in radial density exponent $m$. 
For each choice of $j$ and $m$, we varied the base density $n_0$ of neutral H in 
the $n=2$ state (the normalization factor for disk density) to find the 
value that made the best fit of the observed H$\alpha$ profile. 
The overall best fit was obtained with Keplerian motion and $m=5.5\pm1.5$ 
({\it solid line} in Fig.~6).  The base density in this case was 
$n_0 = 9 \times 10^{-4}$ cm$^{-3}$, 
which corresponds to an electron density of $N_e \approx 2 \times 10^{8}$ cm$^{-3}$. 
This density distribution is very peaked towards the inner radii, but 
models with lower values of $m$ (see the $m=2$ case, {\it dashed line} in Fig.~6)
have worse agreement in the outer line wings (formed close to the star). 
None of the models made assuming angular momentum conservation ($j=1$) 
produced satisfactory fits (see the $m=6$ case, {\it dotted line} in Fig.~6).
The disk rotational velocities are smaller in this case, and the emission 
is redistributed to wavelengths closer to line center, resulting in a worse
match of the emission observed in the line wings. 

This simple, axisymmetric disk model predicts that the out-of-eclipse 
H$\alpha$ profiles will always look like the model profile in Figure~6, 
so the model cannot account for the orbital and temporal variations 
seen in Figure~3.
Nevertheless, the model can be used to predict the kinds of variations
we should observe as the disk is progressively blocked by the secondary 
star around primary eclipse.   We calculated model profiles for the 
eclipse phases by determining which disk elements are occulted by 
the secondary at any instant.   We checked whether or not 
the gravitational potential along the
line of sight from each area element ever attains a value indicating 
a position inside the Roche-filling secondary.   If so, the area 
element was assumed to be occulted by the secondary and was omitted from 
the flux integration.   The predicted eclipse model profiles are plotted 
together with the observed ones in Figures 4 and 5.   The model sequences
show that the approaching, blue-shifted disk contributions are mainly 
occulted at phase $-0.04$, and as the secondary moves it first reveals 
the low velocity blue-shifted emission from the outer disk and 
later emission from the rapidly moving gas close to the star emerges.   
The reverse sequence is seen in the red peak where the high velocity gas close to 
the star is the first to disappear.  Some of these general trends are 
found in the observed spectral sequence.  The predicted changes in the 
extent of the red wing of the emission profile with the advance of the 
eclipse find good agreement with the observations, which supports
the assumption of Keplerian motion in the disk.  However, there are significant
discrepancies beyond the stronger central absorption mentioned above (\S5). 
Since there is strong evidence of temporal variability in the disk (Fig.~3), 
it is possible that the fit of the time-averaged H$\alpha$ profile 
is based on densities that are different from those that actually 
existed at that the times of these two particular sequences.  
It is also possible that there exist asymmetries in the disk 
that could help explain the discrepancies, and in the next
section we use the observed profiles to reconstruct the 
disk surface density distribution. 


\section{Reconstruction Map of the Disk Surface Density}  

The existence of phase-related variations in the H$\alpha$ profile 
that cannot be explained in the context of an axisymmetric disk model
indicates that the disk in RY~Per may have additional non-axisymmetric 
structure (when averaged over many time samples of potentially 
complex structure).   Hydrodynamical simulations of gas flows in 
Algol binaries by \citet{ric98} demonstrate that complicated density 
patterns can develop that produce long standing and large scale patterns 
as well as rapidly changing filamentary structures.   Here we present 
a reconstruction of the time-averaged disk surface density that is based 
upon the orbital variations in the H$\alpha$ emission profiles.  

\citet{ric99} discuss the utility of Doppler tomography 
reconstructions to study the circumstellar matter in Algol binaries. 
The basic assumption in this technique is that circumstellar gas is 
visible in all the spectral observations and that the observed emission 
profile at any particular binary orientation is simply the sum of a 
projection of the vector components of the emitting elements in a 
velocity distribution of gas concentrated towards the orbital plane.  
The method generally involves subtracting a photospheric profile 
appropriate to the mass gainer and then calculating the Doppler 
tomogram of the velocity distribution using various numerical techniques.
\citet{ric99} and \citet{ric01} use the distribution of the emitting 
gas in the velocity tomogram to investigate the spatial distribution 
of circumstellar gas based upon the velocity properties associated 
with the gas stream and disk. 

We show in Figure~7 this kind of Doppler tomogram for the H$\alpha$ 
observations of RY~Per.  This image was made by subtracting the 
photospheric profile shown in Figure~6 from the secondary subtracted 
H$\alpha$ profiles for out-of-eclipse orbital phases 
(Fig.~3), and then performing a back projection tomographic 
reconstruction using the algorithm described by \citet{tha01}.  
The resulting tomogram is shown in the velocity reference frame of 
the primary star where $V_x$ is the velocity component along the 
axis from the primary to the secondary and $V_y$ is the orthogonal 
velocity component in the orbital plane in the direction of orbital motion.
The model line profile corresponding to the tomogram is calculated by  
a projection through this image from a particular orientation (for 
example, the profile for phase 0.00 is a projection from the top of the 
figure to the bottom, while that for phase 0.25 is from the right to 
the left hand side).   We also show in Figure~7 the velocity locations 
of the principal components in the binary that could influence the 
distribution of circumstellar gas.  The large solid-line circle indicates
the velocity limits of the projected rotational velocity of the primary 
while the smaller solid-line figure shows the velocity limits of the Roche-filling
secondary (made assuming synchronous rotation of the secondary).  
The solid-line arc originating at the inner L1 point shows the velocity 
trajectory of a ballistic gas stream which is terminated where the stream 
impacts the primary \citep{lub75}.   The outer long-dashed circle shows the
location of disk Keplerian motion at the primary's surface while the inner
short-dashed circle shows the Keplerian motion at the outer disk edge 
(corresponding to the Roche limit of the primary).  

\begin{figure}
\plotone{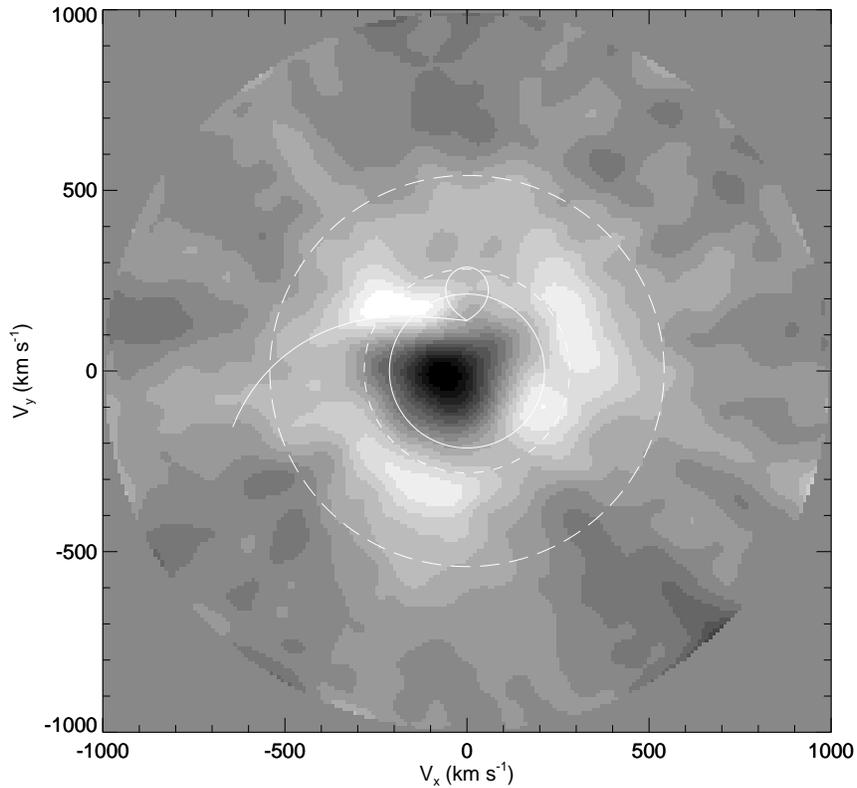}
\caption{A velocity tomogram reconstructed from the out-of-eclipse H$\alpha$ profiles 
of RY~Per (after subtraction of an assumed photospheric component). The 
image displays the net flux in a $201\times201$ velocity grid at increments of
10 km~s$^{-1}$, and the flux is portrayed in a grayscale image ranging from 
$-0.0034$ ({\it black}) to $+0.0029$ ({\it white}) based upon a continuum flux of unity.  
The model profile derived from this image corresponds to a projection 
through the image from a given orientation (from the top for orbital phase 0.00,
from the right for phase 0.25, from the bottom for phase 0.50, and 
from the left for phase 0.75).  The various white lines correspond to 
the velocity locations of binary components discussed in the text.} 
\label{fig7}
\end{figure}

\placefigure{fig7}     

The tomogram image in Figure~7 presents a number of problems that 
challenge the assumptions underlying its calculation.   The 
image is dominated by a dark central absorption that actually 
reaches a minimum flux value below zero (in conflict with 
the premise that the profile is the result of a sum of flux emitting 
elements).  This results from the fact that at the high inclination 
of the RY~Per system part of the disk appears projected against the 
photosphere of the primary and causes absorption of the photospheric flux. 
The absorption properties of the disk are ignored in a straight forward 
application of tomography.  The tomogram image appears to indicate that 
most of the emission originates at velocities associated with the outer disk,
but this conflicts with our result from \S6 that indicated a disk density 
concentration peaked towards the inner part of the disk.   This discrepancy 
is partially due to the optically thick nature of the emission in the inner
disk where the total flux is set by the area of emitting surface and 
the assumed source function (and is insensitive to the actual density).
The occultation of the disk by the primary star will be especially 
important for flux contributions from the inner disk
(for example, we see only half the flux from the innermost disk radius 
at any instant), but this occultation 
of the disk by the star is also neglected in simple tomography.  This is 
not a significant issue for the application of tomography to studies 
of cataclysmic variables where the central white dwarf is small compared 
to the disk dimensions, but it is important in Algol binaries like RY~Per 
where the disk and star sizes are comparable.  

Given the difficulties in interpreting a simple tomogram of the H$\alpha$
emission profiles, we decided to develop a new method to investigate 
the distribution of circumstellar gas in RY~Per.  Our approach uses the 
analytical line formation method outlined in \S6, which accounts for 
geometrical occultation and optical depth effects, together with a 
correction scheme based upon revisions to the assumed disk  
surface density distribution, $\Sigma$ (see \citet{vri99}).   
The goal of the algorithm is to compute the surface density of the disk 
over a grid of radial and azimuthal elements based upon the observed 
profiles over the radial velocity range from $-1000$ to $+1000$ km~s$^{-1}$ 
in all 49 of our spectra.   A coarser grid of 11 radial and 60 azimuthal 
zones was used to describe the disk in the calculation.   We begin by 
setting $\Sigma$ in each zone to the best-fit value determined from 
our $j=0.5$ and $m=5.5$ axisymmetric disk model from \S6.   We then 
consider how a change in surface density affects the flux from a given 
disk element, 
\begin{equation}
{{dF_\lambda}\over{d\Sigma}} = A~(S^L - P_\lambda)~ e^{-\tau_\lambda}~{{\tau_\lambda}\over\Sigma}. 
\end{equation}
In optically thin emitting elements this derivative will only be 
significantly different from zero close to the wavelength corresponding 
to the Doppler shift of the disk element.   On the other hand, in optically 
thick elements the derivative is small near line center and significant 
flux increases can only occur in the ``damping'' wings of the profile 
where $\tau_\lambda \approx 1$.   Thus, to correct the surface density of a
specific disk element we need to compare the observed spectrum $O_\lambda$ with our 
current calculated spectrum $C_\lambda$ at those few wavelengths where the 
absolute value of the derivative attains a maximum.   We selected only those 
wavelength points where the absolute value of the derivative is greater than
$90\%$ of its maximum value.   Such wavelength points are assigned a 
weight $W_\lambda = 1$  while all the other wavelength points have $W_\lambda = 0$.
We then made the simple assumption that the observed deviation can be attributed 
entirely to a surface density correction at that particular element, 
\begin{equation}
{{\triangle\Sigma}\over\Sigma} =
 \int {{\gamma~(O_\lambda - C_\lambda)} \over {A~(S^L - P_\lambda)~ e^{-\tau_\lambda}~\tau_\lambda}} 
 ~W_\lambda ~d\lambda ~/~ \int W_\lambda ~d\lambda .
\end{equation}
In fact there may be many disk elements that have the correct Doppler shift to 
also contribute to the correction $O_\lambda - C_\lambda$ at any specific wavelength, 
so we introduce a gain factor $\gamma$ that is close to zero to make the corrections 
slowly and iteratively.   The algorithm progresses through each spectrum available 
and determines a surface density correction if the element is visible at that 
orbital phase (i.e., free of occultation by the primary and secondary).
Then a mean correction is found from the average of all the available spectra 
and the surface density is revised accordingly.  Since large deviations 
in $\Sigma$ between neighboring disk elements are unphysical, we 
perform a boxcar smoothing in $\log \Sigma$ over 3 adjacent elements 
at this stage.  The entire algorithm is then 
iterated to find new corrections until the root mean square of the 
whole set of $O_\lambda - C_\lambda$ differences reaches a minimum.

We performed a number of tests to determine how well the algorithm 
could reconstruct an arbitrarily defined disk surface distribution 
using a set of model spectra with the same orbital phase distribution and 
instrumental broadening as our observed set.   We found that disk 
azimuthal perturbations of the form $\cos n\phi$ where $n<4$ were successfully 
reconstructed with a typical deviation of $\approx 10\%$ from the 
predefined density over most of the disk.   We also performed tests 
where the radial density distribution was altered in the inner and/or 
outer portions of the disk.  These were also successful except in the 
inner optically thick regions where the 
reconstructed surface densities never departed significantly from 
their initial default values.  This is due to the fact that flux from 
such optically thick elements is relatively insensitive to the density. 
We also found that model density distributions with small regions 
of differing density (either enhanced or reduced) were not reconstructed 
correctly if the affected surface area was less than $10\%$
of the total disk area (such density perturbations appear 
broader and have lower contrast in the reconstructions).  
This stems from the fact that many elements can contribute to a flux 
correction at a given wavelength since many elements share a similar 
radial velocity, and consequently any small deviation seen in the 
profiles tends to be redistributed over many area elements in this kind of reconstruction. 
Thus, we believe the algorithm can succeed in reconstructing large scale 
disk features, but we cannot rely upon it to reproduce fine structure 
or to find accurate density perturbations in optically thick zones. 

We show in Figure~8 the resulting surface density map (in a logarithmic
scaling) based upon all of the observed H$\alpha$ profiles.   
The calculation was made with 9 iterations and a gain $\gamma = 0.02$. 
The spatial dimensions of the disk grid are shown in relation to the system components
in a view from above the orbital plane, with the secondary on the right 
and the primary on the left in the middle of the disk image.   
The dashed line indicates the primary's Roche limit in the orbital plane, 
and the plus sign marks the center of mass.   The dotted line represents
the ballistic trajectory of a gas stream originating at the inner 
Lagrangian point \citep{lub75}.  The disk surface density in this map appears to 
have a somewhat elongated distribution aligned approximately with the axis 
joining the stars, i.e., the surface density is lower in the directions orthogonal 
to this axis (in the vicinity of $\phi = 90^\circ$ and 
$270^\circ$).   Figure~9 plots the average radial decline in disk density
for quadrants centered on $\phi = 90^\circ$ and $270^\circ$
compared with the average for quadrants centered on $\phi = 0^\circ$ and $180^\circ$.
The disk density appears to be approximately $4\times$ lower in the regions 
orthogonal to the axis.   Figure~8 also indicates that the disk may have a local 
density enhancement in the region where the gas stream approaches the primary.

\begin{figure}
\rotatebox{90}{
\epsscale{0.7}
\plotone{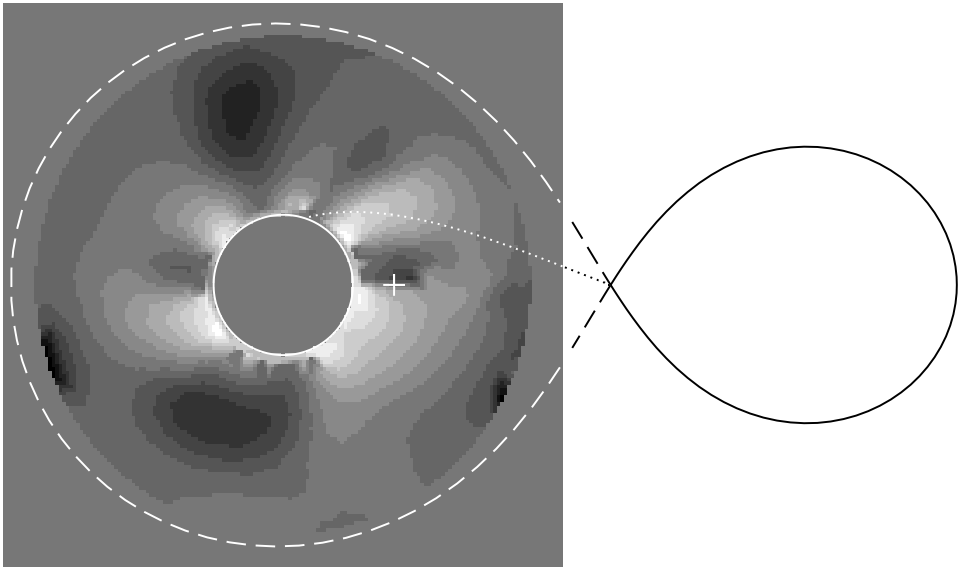}}
\caption{A 16 level grayscale depiction of the logarithm of the disk surface
density in a spatial diagram as viewed from above the orbital plane. 
The Roche-filling secondary is shown in outline on the right
while the primary star is located at the center of the disk.  
The dashed line indicates the Roche limit of the primary, and 
the dotted line shows the ballistic trajectory of the gas stream
from the secondary star.} 
\label{fig8}
\end{figure}

\placefigure{fig8}     

\begin{figure}
\rotatebox{90}{
\epsscale{0.7}
\plotone{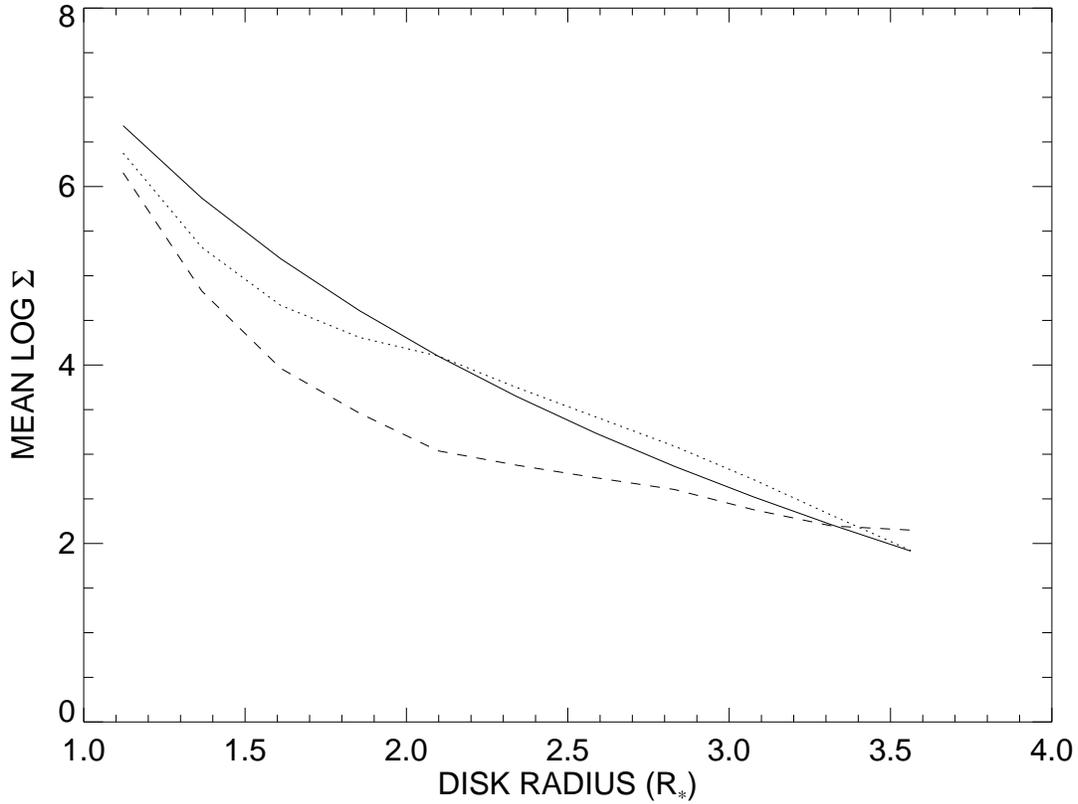}}
\caption{The logarithm (base 10) of the surface density as a function of radial 
distance in the disk based upon the axisymmetric model ({\it solid line}). 
The radial dependence is also shown for the reconstructed surface density map 
for the average of two quadrants centered on the axis 
($\phi = 0^\circ$ and $180^\circ$, {\it dotted line})
and the average of two quadrants orthogonal to the axis 
(centered on $\phi = 90^\circ$ and $270^\circ$, {\it dashed line}).} 
\label{fig9}
\end{figure}

\placefigure{fig9}     

\citet{ric98} made hydrodynamical simulations of the gas flows 
of two Algol binary systems, $\beta$~Per and TT~Hya.  The systems differ 
in where the gas stream encounters the primary: the stream hits the 
primary in $\beta$~Per near an azimuth of $\phi \approx 60^\circ$ 
while it misses the limb of the primary altogether for TT~Hya. 
The stream geometry in RY~Per lies between these cases but we do 
expect that a stream -- star impact occurs (Fig.~8).   
The best comparison can be made with their simulation for $\beta$~Per
at the conclusion of a long time sequence (see Fig.~6 in \citet{ric98}). 
Their results show that some of the incoming gas stream is deflected into
plumes that extend in the direction of $\phi=180^\circ$, and some of
this material passes close to the primary on a return trajectory to 
occupy the region around $\phi=0^\circ$.  We suspect that the
elongated surface density distribution we find for RY~Per (Fig.~8) 
indicates that similar kinds of flows are occurring in its disk 
(at least in a time-averaged sense based upon results from spectra 
taken on many dates). 

The hydrodynamical simulations of \citet{ric98} are also important
in demonstrating some of the limitations of our disk mapping 
techniques.  \citet{ric98} find that the gas temperature 
varies significantly in the circumstellar material, attaining 
peak values near the impact site and in diffuse regions between 
dense gas filaments that surround the primary 
(the existence of a hot disk component is supported by 
UV observations of highly ionized emission lines; \citet{pet01}). 
Our model assumes an isothermal disk, so it is possible that some disk elements in 
our model (for example, those near the hot impact site) are assigned too low a 
surface density because H$\alpha$ emissivity 
declines with increasing temperature \citep{ric98}.   
However, the hydrodynamical simulations suggest that, with the exception 
of the always hot impact zone, most disk regions will experience similarly 
diverse temperature fluctuations, so that a time-averaged map like ours 
should not be too adversely affected by the assumption of constant temperature. 
The hydrodynamical simulations also show that the gas flow velocities 
can depart significantly from Keplerian motion, particularly in regions 
close to the star.   Our model will suffer by assigning an incorrect 
velocity to certain disk elements, but since the main effect of gas close 
to the star is to create increased absorption of the photospheric flux 
where such elements are moving nearly tangentially to the line of sight, 
we doubt that non-Keplerian flows will seriously alter the appearance 
of the surface density map. 

The inclusion of the disk asymmetries in our model line profiles 
decreases the root mean square of the $O_\lambda - C_\lambda$ 
residuals from 0.035 of the continuum flux in the 
axisymmetric case (\S6) to 0.031 in the converged model.   
Taken at face value this is only 
a modest improvement, but recall that most of this difference 
is due to temporal variations which are probably larger than the 
orbital phase-related variations (Fig.~3).   The simulated spectra 
based upon the density map in Figure~8 are shown as dashed lines 
in Figures 3 -- 5, and on the whole these model profiles provide 
a better match of the observations than those based upon the 
axisymmetric model.   This is especially true in the eclipse 
sequences; for example, the blue wing variations in Figure~4 
are much more closely reproduced when the disk density is 
lower in regions orthogonal to the axis joining the stars. 
Thus, we conclude that the disk density structures indicated 
in the reconstructed map are required to model adequately 
the observed H$\alpha$ variations. 


\section{Conclusions}                               

The H$\alpha$ observations of RY~Per indicate that the mass
gainer in this Algol-type binary is surrounded by a time variable 
accretion disk.   The H$\alpha$ profile usually appears as a 
double-peaked emission line with a strong central absorption.   
This same profile morphology appears in the Be shell stars, and 
we fit the mean H$\alpha$ profile of RY~Per using  
the line formation method that \citet{hum00} have applied to 
Be stars.   We can match the observed and model profiles
if the accretion disk has a steep radial decline 
in surface density and experiences Keplerian motion.   
The calculation demonstrates the importance of accounting 
for disk absorption in the analysis of the emission lines.  
The presence of orbital phase-related variations in the H$\alpha$ 
profile implies that the disk is structured, and we developed 
a new method of tomographic reconstruction to derive a 
map of the disk surface density (based upon the spatial 
and velocity relationship for Keplerian motion).   
The density map shows similarities to the results of hydrodynamical 
simulations of gas flows in Algols \citep{ric98} 
and suggests that the incoming gas stream is deflected 
at the impact site into elongated trajectories that create 
density enhancements along the axis joining the stars. 

Our interest in RY~Per was sparked by its rapid rotation (\S4)
which is probably the result of the accretion of gas
with high angular momentum.   The dimensions of the RY~Per 
system are such that the gas stream strikes the star very close 
to its trailing edge (Fig.~8), and this is clearly a favorable 
situation to impart angular momentum from the gas stream to 
the primary.   Algol binaries become more widely separated as 
mass transfer continues \citep{hil01}, so that in the past 
when the components of RY~Per were closer the gas stream would 
have hit the primary closer to its center, a less favorable 
geometry to gain angular momentum.   We can calculate how 
the angular momentum gain varies as the system mass ratio 
changes by considering the angle $\psi$ between the surface normal
and the gas stream velocity vector $\vec{V}$ at the point of impact. 

Figure~10 shows a sample calculation of the magnitude of 
$\vec{V}\times\vec{R} = V R \sin\psi$ based upon a conservative 
mass transfer evolutionary path for RY~Per.  We constructed 
a series of gas stream trajectories for a grid of mass ratios 
$q=M_S/M_P$ \citep{lub75}.  
In each case the system semi-major axis was computed according to 
\begin{equation}
{a\over{a_1}} = {1\over{16}}~ {{(1+q)^4}\over{q^2}}
\end{equation}
where $a_1$ was the closest separation at the time when 
the stars had the same mass ($q=1$). 
The radius of the mass gainer was estimated from its mass 
according to the mass -- radius relation for main sequence stars, 
$R\propto M^{0.64}$.  
The point of impact and the value of $\vec{V}\times\vec{R}$ were 
derived for each test value of $q$.
We found that for mass ratios $q<0.22$ the stream missed the 
star altogether.  We assume in these cases that the gas will end 
up in an accretion disk and will eventually reach the stellar surface 
with the Keplerian velocity.    Figure~10 shows that 
as the system evolves to lower values of $q$ the mass gainer 
receives more and more angular momentum per gram of gas accreted. 
It is interesting to note that the current position of RY~Per in
this diagram (indicated by the solid dot) is near the peak 
of the angular momentum transfer efficiency.  

\begin{figure}
\rotatebox{90}{
\epsscale{0.7}
\plotone{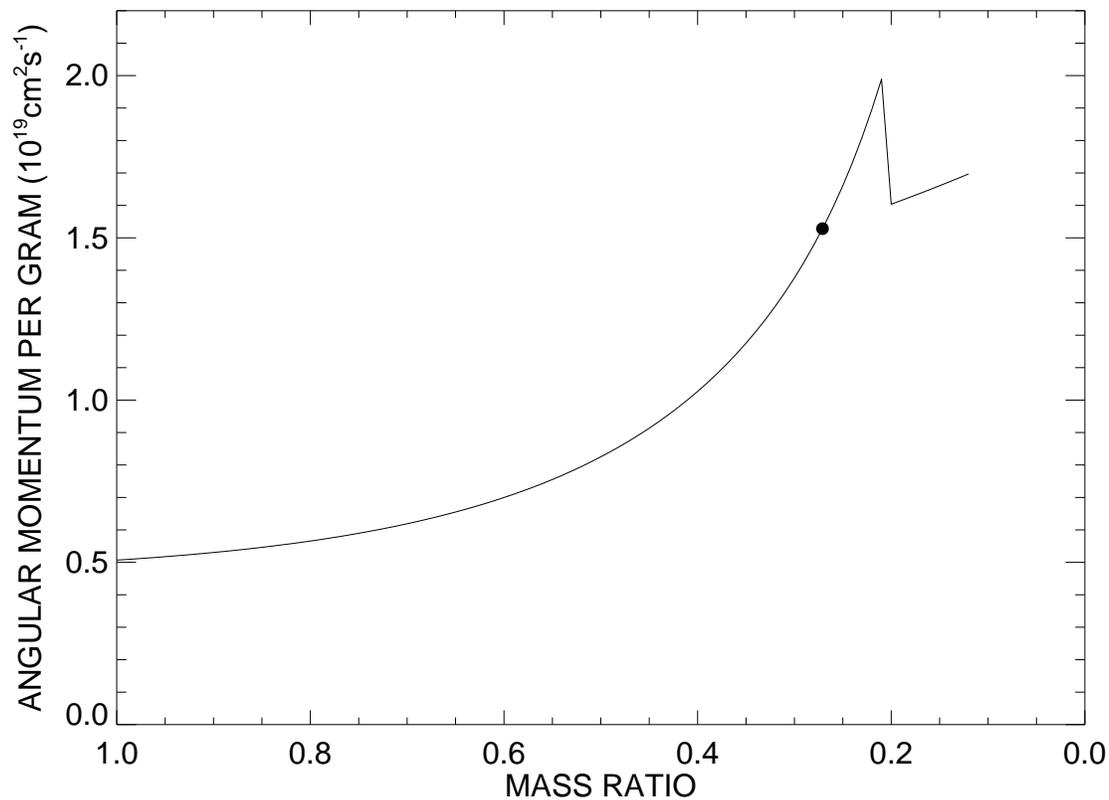}}
\caption{The evolution of angular momentum transfer efficiency for 
a conservative mass transfer scenario for RY~Per.  The curve shows 
the amplitude of $\vec{V}\times\vec{R}$, the vector component of
angular momentum from the gas stream at the point of impact, 
as function of $q=M_S/M_P$ (which decreases with time).  The solid 
dot shows the current state of RY~Per.  The discontinuity near 
$q=0.2$ marks the boundary where the gas stream misses the star and 
angular momentum is accreted according to the Keplerian velocity
at the stellar surface.} 
\label{fig10}
\end{figure}

\placefigure{fig10}     

The angular momentum efficiency variation suggests that 
the spin up of the mass gainer may become more pronounced as 
systems approach the peak of the efficiency curve.  
Other factors such as the mass transfer rate and tidal 
and magnetic breaking will also play important roles in 
the effective spin up of the gainer star.   Nevertheless, 
it is worthwhile considering whether other systems like
RY~Per also show evidence of spin up when they reach the 
efficiency maximum.   Figure~11 shows the location of RY~Per 
({\it solid circle}) in the $r-q$ diagram in which the fractional 
radius of the gainer is plotted against the system mass ratio.  
The solid line shows the evolutionary path followed by RY~Per
as the system evolves from large to small $q$ and the semimajor
axis increases.   
The dotted curve labelled by $\bar{\omega}_{\rm min}$ shows the 
minimum distance between the gas stream and center of the gainer
\citep{lub75}, and once the gainer's radius is smaller than this
value the gas stream will miss the star and form a disk 
(the intersection of the curves corresponds to the discontinuity
in Fig.~10).   Algol binaries found below $\bar{\omega}_{\rm min}$ 
generally have well developed, permanent accretion disks \citep{pet01}. 
The dashed curve marked by $\bar{\omega}_{\rm d}$ is the fractional 
outer radius of a disk in which the orbital velocity at the 
stream -- disk intersection matches the vector component of 
the stream in the same direction \citep{lub75}.  Most dense accretion 
disks can form only if the stellar radius is smaller than $\bar{\omega}_{\rm d}$, 
and Algols above this line show no evidence of disks \citep{pet01}.
RY~Per falls in the region between the curves, and other systems 
in this same region have variable disks \citep{pet01}. 

\begin{figure}
\rotatebox{90}{
\epsscale{0.7}
\plotone{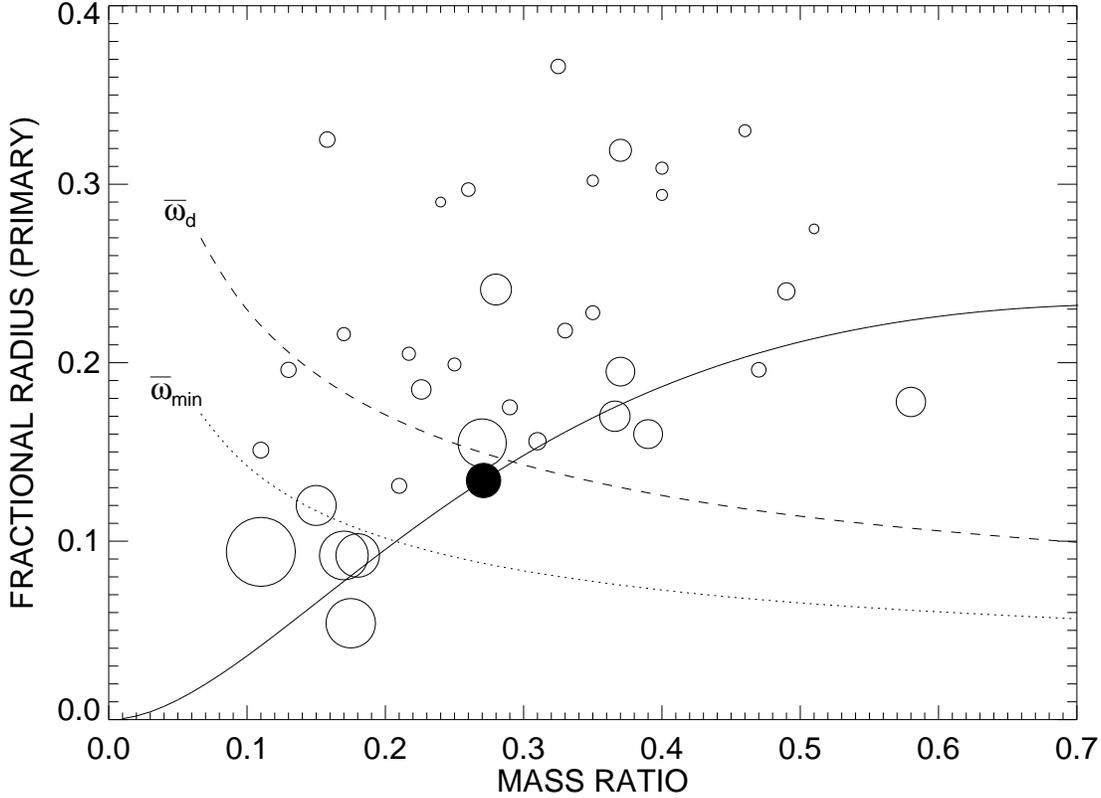}}
\caption{An $r,q$ diagram showing the location of 
the mass gainers in RY~Per ({\it solid circle})
and other Algol binaries \citep{van90} ({\it open circles}).  The area of 
each circle is proportional to the ratio of the gainer's angular velocity 
compared to the synchronous rate.   The solid line shows the evolutionary 
path of RY~Per for conservative mass transfer.  
The dotted curve labelled by $\bar{\omega}_{\rm min}$ shows the 
minimum distance between the gas stream and center of the gainer, 
and the dashed curve marked by $\bar{\omega}_{\rm d}$ is the fractional 
outer radius of a disk in which the orbital velocity at the 
stream -- disk intersection matches the vector component of 
the stream in the same direction \citep{lub75}.  
} 
\label{fig11}
\end{figure}

\placefigure{fig11}     

We also show in Figure~11 the positions of gainers in 34 other 
Algol binaries from the study of \citet{van90} (see their Table 6). 
\citet{van90} list the ratio of primary's angular spin velocity 
to the orbital angular velocity, and the open circular area plotted 
for each system in Figure~11 is proportional to this ratio
(rapid rotators have larger symbols).  There appears to be a 
trend for the rapid rotators to appear in the lower left portion 
of this diagram, which corresponds to the evolutionary stage 
near the peak of the angular momentum transfer efficiency (Fig.~10).  
This suggests that gainers in most Algols are significantly spun up 
once they reach the stage where the gas stream strikes the star 
nearly tangentially (the epoch in which we now find RY~Per). 

The example of RY~Per demonstrates clearly how massive stars 
can be spun by mass transfer in close binaries.  Such systems will 
increase their separation as mass transfer progresses making it more 
and more difficult for tidal forces to slow the rotation of the 
gainer, and consequently the gainer may remain a rapid rotator 
for most of its main sequence lifetime (barring further interaction 
with the secondary).   Since many stars are members of close binaries, 
these mass and angular momentum transfer processes could plausibly 
account for a significant fraction of the population of rapidly rotating 
massive stars \citep{pol91,vbv97,gie01}.  


\acknowledgments

We thank the staff of KPNO 
for their assistance in making these observations possible. 
We are grateful for comments on this work by Walter Van Hamme 
and an anonymous referee.  Financial support was provided
by the National Science Foundation through grant AST$-$0205297 (DRG).
Institutional support has been provided from the GSU College
of Arts and Sciences and from the Research Program Enhancement
fund of the Board of Regents of the University System of Georgia,
administered through the GSU Office of the Vice President
for Research.  This research made use of
the Multimission Archive at the Space Telescope Science Institute (MAST)
and NASA's Astrophysics Data System Bibliographic Service. 




\clearpage


\begin{deluxetable}{llrcccc}
\tablewidth{0pc}
\tablecaption{KPNO Coude Feed Observing Runs\label{tab1}}
\tablehead{
\colhead{UT Dates} &
\colhead{G/$\lambda_b$/O\tablenotemark{a}} &
\colhead{Filter} &
\colhead{$\lambda / \Delta\lambda$} &
\colhead{Range (\AA)} }
\startdata
 1999 Oct 12 -- 18 \dotfill & 316/~7500/1  & GG495  &  ~4100 & 5400 -- 6736 \\
 1999 Nov 09 -- 15 \dotfill & 316/~7500/1  & GG495  &  ~4400 & 5545 -- 6881 \\
 1999 Dec 04 -- 09 \dotfill & 632/12000/2  & OG550  &  31000 & 6456 -- 6774 \\
 2000 Oct 02 -- 04 \dotfill & 316/12000/2  & OG550  &  ~9500 & 6440 -- 7105 \\
\enddata
\tablenotetext{a}{Grating grooves mm$^{-1}$; blaze wavelength (\AA ); order.}
\end{deluxetable}
\clearpage



\begin{deluxetable}{lccc}
\small
\tablewidth{0pc}
\tablecaption{Radial Velocities for the Secondary \label{tab2}}
\tablehead{
\colhead{HJD}           &
\colhead{Spectroscopic} &
\colhead{$V_r$}         &
\colhead{$O-C$}         \\
\colhead{(2,400,000$+$)}  &
\colhead{Phase}         &
\colhead{(km s$^{-1}$)} &
\colhead{(km s$^{-1}$)} }
\startdata
 51463.715 \dotfill &  0.526 & 
\phn         $ -23.1$ &\phn\phs $   0.3$ \\
 51463.737 \dotfill &  0.529 & 
\phn         $ -17.0$ &\phn\phs $   3.1$ \\
 51463.761 \dotfill &  0.533 & 
\phn         $ -15.5$ &\phn\phs $   1.0$ \\
 51463.783 \dotfill &  0.536 & 
\phn         $ -12.1$ &\phn\phs $   1.2$ \\
 51463.804 \dotfill &  0.539 & 
\phn\phn     $  -9.4$ &\phn\phs $   0.8$ \\
 51463.828 \dotfill &  0.543 & 
\phn\phn     $  -4.6$ &\phn\phs $   1.9$ \\
 51463.850 \dotfill &  0.546 & 
\phn\phn     $  -2.0$ &\phn\phs $   1.3$ \\
 51463.871 \dotfill &  0.549 & 
\phn\phn     $  -1.8$ &\phn     $  -1.6$ \\
 51463.895 \dotfill &  0.552 & 
\phn\phn\phs $   3.3$ &\phn     $  -0.2$ \\
 51463.916 \dotfill &  0.555 & 
\phn\phn\phs $   4.9$ &\phn     $  -1.7$ \\
 51463.937 \dotfill &  0.558 & 
\phn\phs     $  10.6$ &\phn\phs $   0.8$ \\
 51463.960 \dotfill &  0.562 & 
\phn\phs     $  14.0$ &\phn\phs $   0.8$ \\
 51463.992 \dotfill &  0.566 & 
\phn\phs     $  16.8$ &\phn     $  -1.1$ \\
 51464.874 \dotfill &  0.695 & 
\phs         $ 132.4$ &\phn\phs $   0.5$ \\
 51465.732 \dotfill &  0.820 & 
\phs         $ 168.9$ &\phn     $  -0.1$ \\
 51465.982 \dotfill &  0.856 & 
\phs         $ 155.2$ &\phn     $  -4.4$ \\
 51466.722 \dotfill &  0.964 & 
\phn\phs     $  81.8$ &\phn\phs $   2.1$ \\
 51466.953 \dotfill &  0.998 & 
\phn\phs     $  41.5$ &\phn     $  -1.7$ \\
 51466.975 \dotfill &  0.001 & 
\phn\phs     $  37.7$ &\phn     $  -1.9$ \\
 51467.745 \dotfill &  0.113 & 
\phn         $ -90.6$ &\phn     $  -2.9$ \\
 51467.766 \dotfill &  0.116 & 
\phn         $ -87.7$ &\phn\phs $   3.1$ \\
 51467.945 \dotfill &  0.142 & 
             $-111.2$ &\phn\phs $   4.3$ \\
 51467.966 \dotfill &  0.145 & 
             $-114.9$ &\phn\phs $   3.2$ \\
 51468.730 \dotfill &  0.257 & 
             $-180.8$ &\phn     $  -3.8$ \\
 51468.909 \dotfill &  0.283 & 
             $-181.7$ &\phn     $  -2.8$ \\
 51468.930 \dotfill &  0.286 & 
             $-180.4$ &\phn     $  -1.7$ \\
 51469.739 \dotfill &  0.404 & 
             $-128.8$ &\phn\phs $   4.0$ \\
 51469.760 \dotfill &  0.407 & 
             $-123.3$ &\phn\phs $   7.3$ \\
 51469.941 \dotfill &  0.433 & 
             $-104.3$ &\phn\phs $   6.2$ \\
 51491.858 \dotfill &  0.626 & 
\phn\phs     $  73.7$ &\phn     $  -3.0$ \\
 51492.802 \dotfill &  0.764 & 
\phs         $ 166.2$ &\phn\phs $   1.3$ \\
 51493.798 \dotfill &  0.909 & 
\phs         $ 130.3$ &\phn\phs $   0.9$ \\
 51494.808 \dotfill &  0.056 & 
\phn         $ -27.1$ &\phn     $  -2.2$ \\
 51494.830 \dotfill &  0.059 & 
\phn         $ -29.9$ &\phn     $  -1.4$ \\
 51495.852 \dotfill &  0.208 & 
             $-161.7$ &\phn     $  -0.4$ \\
 51496.844 \dotfill &  0.353 & 
             $-172.0$ &\phn     $  -9.8$ \\
 51497.811 \dotfill &  0.494 & 
\phn         $ -56.8$ &\phn     $  -1.2$ \\
 51516.669 \dotfill &  0.241 & 
             $-170.5$ &\phn\phs $   3.1$ \\
 51517.664 \dotfill &  0.386 & 
             $-143.3$ &\phn\phs $   0.9$ \\
 51521.681 \dotfill &  0.971 & 
\phn\phs     $  74.1$ &\phn\phs $   2.0$ \\
 51819.777 \dotfill &  0.403 &
             $-134.9$ &\phn     $  -1.7$ \\
 51820.677 \dotfill &  0.534 & 
\phn         $ -15.0$ &\phn    $   -0.1$ \\
 51820.714 \dotfill &  0.540 & 
\phn\phn     $  -9.2$ &\phn\phs $   0.3$ \\
 51820.763 \dotfill &  0.547 & 
\phn\phn     $  -4.2$ &\phn     $  -2.1$ \\
 51820.806 \dotfill &  0.553 & 
\phn\phn\phs $   3.8$ &\phn     $  -0.4$ \\
 51820.852 \dotfill &  0.560 & 
\phn\phs     $  11.3$ &\phn\phs $   0.2$ \\
 51820.896 \dotfill &  0.566 & 
\phn\phs     $  12.8$ &\phn     $  -4.9$ \\
 51820.962 \dotfill &  0.576 & 
\phn\phs     $  20.2$ &\phn     $  -7.2$ \\
 51821.842 \dotfill &  0.704 & 
\phs         $ 145.2$ &\phn\phs $   7.4$ \\
\enddata
\end{deluxetable}
\clearpage



\begin{deluxetable}{lcccl}
\small
\tablewidth{0pc}
\tablecaption{Radial Velocities for the Primary \label{tab3}}
\tablehead{
\colhead{HJD}           &
\colhead{Spectroscopic} &
\colhead{$V_r$}         &
\colhead{$O-C$}         \\
\colhead{(2,400,000$+$)}  &
\colhead{Phase}         &
\colhead{(km s$^{-1}$)} &
\colhead{(km s$^{-1}$)} }
\startdata
 48155.075 \dotfill &  0.441 & 
\phn\phs     $  40.8$ &\phn\phs $   0.4$ \\
 49650.939 \dotfill &  0.383 & 
\phn\phs     $  55.8$ &\phn\phs $   3.9$ \\
 49652.926 \dotfill &  0.672 & 
\phn         $ -30.3$ &         $ -11.0$ \\
 49653.947 \dotfill &  0.821 & 
\phn         $ -31.5$ &\phn\phs $   2.1$ \\
 49654.935 \dotfill &  0.965 & 
\phn\phn\phs $   1.5$ &\phs     $  10.5$ \\
 49655.941 \dotfill &  0.112 & 
\phn\phs     $  25.8$ &\phn     $  -9.9$ \\
 49656.967 \dotfill &  0.261 & 
\phn\phs     $  56.6$ &\phn     $  -3.8$ \\
 49659.690 \dotfill &  0.658 & 
\phn         $ -16.0$ &\phn\phs $   0.1$ \\
 49671.873  \dotfill &  0.433 & 
\phn\phs     $  49.8$ &\phn\phs $   7.6$ \\
\enddata
\end{deluxetable}

\clearpage



\begin{deluxetable}{lcc}
\tablewidth{0pc}
\tablecaption{Orbital Elements\label{tab4}}
\tablehead{
\colhead{Element}                & \colhead{Popper (1989)}    & \colhead{This Work}}
\startdata
$P$ (d)                 \dotfill & 6.863569\tablenotemark{a}  & 6.863569\tablenotemark{a} \\
$T_P$ (HJD -- 2,400,000)\dotfill & \nodata                    & $51467.15 \pm 0.10$ \\
$T_S$ (HJD -- 2,400,000)\dotfill & \nodata                    & $51466.97 \pm 0.13$ \\
$e$                     \dotfill & 0.0\tablenotemark{b}       & $0.036 \pm 0.005$ \\
$\omega_P$ (deg)        \dotfill & \nodata                    & 255\tablenotemark{b} \\
$\omega_S$ (deg)        \dotfill & \nodata                    & $75 \pm 7$ \\
$K_P$ (km s$^{-1}$)     \dotfill & $50.1 \pm 2.1$             & $47.3 \pm 3.9$ \\
$K_S$ (km s$^{-1}$)     \dotfill & $175.8 \pm 1.2$            & $174.5 \pm 0.9$ \\
$V_P$ (km s$^{-1}$)     \dotfill & $-11.6 \pm 1.5$            & $13.8 \pm 2.9$ \\
$V_S$ (km s$^{-1}$)     \dotfill & $-6 \pm 2$                 & $-6.0 \pm 0.6$ \\
rms$_P$ (km s$^{-1}$)   \dotfill & 8.8                        & 8.4 \\
rms$_S$ (km s$^{-1}$)   \dotfill & 5.7                        & 3.4 \\
$M_P$ ($M_\odot$)       \dotfill & \nodata                    & $6.24 \pm 0.24$ \\
$M_S$ ($M_\odot$)       \dotfill & \nodata                    & $1.69 \pm 0.23$ \\
$a$ ($R_\odot$)         \dotfill & \nodata                    & $30.3 \pm 0.6$ \\
\enddata
\tablenotetext{a}{From the light curve analysis of Van Hamme \& Wilson (1986).}
\tablenotetext{b}{Fixed.}
\end{deluxetable}
\clearpage



\clearpage

\end{document}